\documentclass{article}
\usepackage{ijcai25}

\usepackage{times}
\usepackage{soul}
\usepackage{url}
\usepackage[hidelinks]{hyperref}
\usepackage[utf8]{inputenc}
\usepackage[small]{caption}
\usepackage{graphicx}
\usepackage{amsmath}
\usepackage{amsthm}
\usepackage{booktabs}
\usepackage{algorithm}
\usepackage{algorithmic}
\usepackage[switch]{lineno}
\usepackage{multirow}
\usepackage{bm}

\usepackage{subcaption}
\usepackage{amsfonts}  

\title{MHANet: Multi-scale Hybrid Attention Network for Auditory Attention Detection}

\author{
Lu Li
\and
Cunhang Fan\thanks{Corresponding author} 
\and
Hongyu Zhang
\and
Jingjing Zhang
\and
Xiaoke Yang
\and
Jian Zhou 
\And
Zhao Lv 
\affiliations
Anhui Province Key Laboratory of Multimodal Cognitive Computation, School of Computer Science and Technology, Anhui University, Hefei, 230601, China
\emails
\{e12314059, e22201103, e22201067, e22201014\}@stu.ahu.edu.cn\\
\{cunhang.fan, Jzhou, kjlz\}@ahu.edu.cn
}

\begin{document}

\maketitle

\begin{abstract}
Auditory attention detection (AAD) aims to detect the target speaker in a multi-talker environment from brain signals, such as electroencephalography (EEG), which has made great progress. However, most AAD methods solely utilize attention mechanisms sequentially and overlook valuable multi-scale contextual information within EEG signals, limiting their ability to capture long-short range spatiotemporal dependencies simultaneously. To address these issues, this paper proposes a multi-scale hybrid attention network (MHANet) for AAD, which consists of the multi-scale hybrid attention (MHA) module and the spatiotemporal convolution (STC) module. Specifically, MHA combines channel attention and multi-scale temporal and global attention mechanisms. This effectively extracts multi-scale temporal patterns within EEG signals and captures long-short range spatiotemporal dependencies simultaneously. To further improve the performance of AAD, STC utilizes temporal and spatial convolutions to aggregate expressive spatiotemporal representations. Experimental results show that the proposed MHANet achieves state-of-the-art performance with fewer trainable parameters across three datasets, 3 times lower than that of the most advanced model. Code is available at: \href{https://github.com/fchest/MHANet}{https://github.com/fchest/MHANet}.
\end{abstract}

\section{Introduction}
In a complex auditory scene where multiple people speak simultaneously, commonly referred to as the cocktail party problem~\cite{party}, humans are capable of directing their auditory attention on one particular speaker. However, individuals with hearing impairments often face significant challenges in identifying and focusing on the attended speaker. Previous neuroscience studies have demonstrated that there exists a connection between brain activity and auditory attention~\cite{5}. The auditory attention detection (AAD) task aims to detect auditory attention from neural activities. Inspired by these findings, researchers have proposed various neurorecording modalities to address this issue, including electroencephalography (EEG)~\cite{6,7}, electrocorticography (ECoG)~\cite{5}, and magnetoencephalography (MEG)~\cite{8,9}. Among these, EEG-based methods stand out as premier solutions due to the non-invasive nature, wearability, and relative affordability of EEG. Therefore, in this paper, we concentrate on utilizing EEG signals for AAD.

According to studies that the cortical responses to an attended speaker are encoded within EEG signals, which correlate with auditory stimulus~\cite{doi:10.1073/pnas.1523357113,BEDNAR2020116283}, the linear methods reconstruct the stimulus from EEG signals to detect the correlation between the reconstructed stimulus and the attended speech envelopes~\cite{ais,dcc}. However, in most real-world scenarios, obtaining clean auditory stimuli is challenging, on which these methods heavily depend. Meantime, the cortical responses have a nonlinear relationship with the acoustic stimuli~\cite{10.7554/eLife.53445}, and these linear methods have difficulty in capturing this complexity, thus leading to a longer decision window. 

Therefore, some studies attempt to employ nonlinear neural networks to directly relate raw EEG signals to the attention detection decision~\cite{Ciccarelli2019,monesi2020lstm,10.7554/eLife.56481}. For instance, the long short term memory (LSTM) layer is used to capture temporal patterns of EEG signals~\cite{monesi2020lstm}. Then, a convolutional neural network (CNN) is proposed to extract the locus of auditory attention~\cite{10.7554/eLife.56481}. Recently, some research has demonstrated that the latent spatial distribution of different EEG channels improves AAD performance. Consequently, some studies project the extracted differential entropy (DE) values on two-dimensional (2D) topological maps~\cite{mbssf,DBPNet,fan2025seeing} or transform the original EEG channels into a 2D spatial topological map~\cite{xu2024densenet}. To capture the dynamic auditory attention activity sensitive to the temporal patterns~\cite{ZIONGOLUMBIC2013980}, the self-attention mechanism, which is proposed in~\cite{vaswani2017attention} and widely used in numerous vision tasks~\cite{RN123}, is also introduced into AAD to learn mutual temporal relationships or spatial distribution features within EEG signals~\cite{stanet,XANet}. However, most methods usually have limited capabilities in comprehensively considering spatiotemporal representations of EEG signals. 

To solve the problem, some studies adopt sequential approaches by applying the spatial attention mechanism followed by the temporal attention mechanism~\cite{stanet}. Nevertheless, EEG signals inherently possess spatiotemporal characteristics, and the separate consideration of spatial and temporal dimensions hinders these models from capturing spatiotemporal dependencies in an integrated manner. Simultaneously, the fusion of representations across multiple temporal and spatiotemporal scales is critical for uncovering latent brain activities related to auditory attention. Unfortunately, most existing models fail to adequately consider multi-scale features of EEG signals, leading to suboptimal feature representation and sensitivity to noise. This oversight significantly restricts their performance of AAD.

To address these issues, this paper proposes a novel multi-scale hybrid attention network for AAD, named MHANet, which effectively captures spatiotemporal features at multiple scales and uncovers the latent dependencies between spatial distribution features and temporal features from a global perspective. Specifically, our model comprises two modules: (1) \textit{Multi-scale Hybrid Attention Module}. This module integrates a multi-scale temporal attention (MTA) block with the channel attention (CA) mechanism to combine expressive multi-scale temporal information and spatial distribution features, enhancing the representation capacity of EEG signals. Additionally, we introduce a multi-scale global attention (MGA) block, which treats EEG signals as Euclidean feature maps to simultaneously capture spatial distribution features and temporal patterns at multiple scales. This allows our model to focus on key channels while capturing long-short range temporal contextual information globally. (2) \textit{Spatiotemporal Convolution Module}. In this module, multi-scale spatiotemporal features are aggregated through spatial and temporal layers, followed by global average pooling, to consolidate comprehensive representations.
We evaluate the AAD performance of our MHANet on three datasets: KUL, DTU, and AVED. The results demonstrate that MHANet achieves state-of-the-art (SOTA) performance across three datasets. The major contributions of this paper are as follows:
\begin{itemize}
    \item A novel AAD network, named MHANet, is proposed in this paper. This architecture combines multi-scale temporal features and spatial distribution features to capture long-short range spatiotemporal dependencies simultaneously.
    \item We introduce MTA to extract temporal information of EEG signals at multiple scales, enabling the construction of comprehensive temporal representations. Moreover, we propose MGA to capture multi-scale spatiotemporal dependencies globally, effectively identifying key channels and temporal patterns within EEG signals. 
    \item The MHANet achieves SOTA decoding accuracy within an extremely short 0.1-second decision window on the KUL dataset, with an accuracy of 95.6\%. It outperforms the best model by 6.4\%. Moreover, our model is highly efficient with only 0.02M parameters, 3 times fewer than the leading model.
\end{itemize}
\begin{figure*}[ht]
    \centering
    \includegraphics[width=1\linewidth]{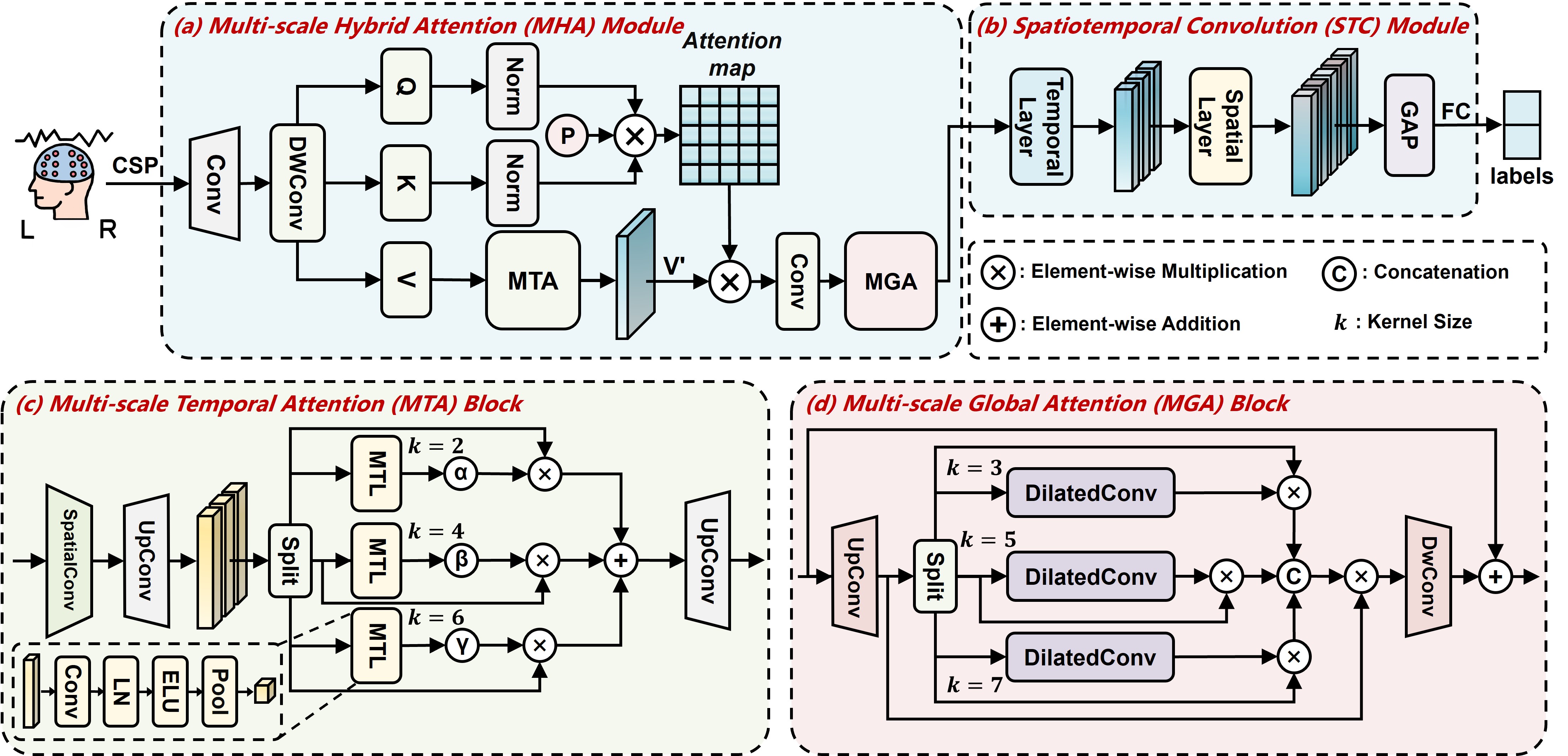}
    \caption{The overview architecture of our MHANet model for AAD, which mainly consists of two modules: (a) multi-scale hybrid attention (MHA) module and (b) spatiotemporal convolution (STC) module.}
    \label{model-figure}
\end{figure*}
\section{The proposed MHANet}
\subsection{Overall Architecture}
The previous AAD models typically apply attention mechanisms sequentially and neglect long-short range dependencies of spatiotemporal features of EEG signals at multiple scales.

To address these issues, we propose MHANet, as shown in Figure~\ref{model-figure}, which consists of MHA and STC. Specifically, MHA includes CA with a MTA and a MGA. Our MHANet extracts key spatial distribution features and long-short range temporal information at multiple scales, thus establishing a strong relationship between them.

Firstly, the processed data \(E\) is fed into MHA to extract comprehensive spatiotemporal features. Then, the output data is aggregated through STC. Finally, a fully connected (FC) layer is applied to generate the final prediction \(p\). 
\begin{equation}
    p = w(STC(MHA(E)) + b 
\end{equation}
where \(w\) and \(b\) are the weight and bias parameters of the FC layer. We apply the cross-entropy loss function to supervise the network's training.

In the following, we introduce the proposed MHA, MTA, MGA and STC.

\subsection{Multi-scale Hybrid Attention Module}

As demonstrated in previous studies~\cite{arvaneh2011optimizing}, EEG signals capture the brain's neuronal electrical activity, which varies over time and reflects activity patterns and connectivity across different brain regions. This temporal and spatial variability makes it possible to analyze the brain's response patterns to auditory stimuli by deeply extracting spatiotemporal features from EEG signals.

However, previous studies solely apply sequential attention mechanisms and focus on isolated spatial or temporal information. To address these shortcomings, we combine a MTA with CA~\cite{9878962} to comprehensively extract multi-scale spatiotemporal features from EEG signals. This architecture ensures that both temporal dynamics and spatial relationships are robustly captured, enhancing the decoding of auditory attention.
 
Firstly, we employ a convolutional layer to increase the channel resolution, enhancing the representation of spatial relationships of EEG signals. Following this, a depth-wise convolution is applied to extract temporal features \(E' \) within each individual channel, focusing on the temporal dynamics within EEG signals. This can be formulated as follows:
\begin{equation}
    E' = DWConv(Conv(E)) \in \mathbb{R}^{3C \times 1 \times T}
\end{equation}
where \(Conv(\cdot)\) represents a \(1 \times 1\) convolutional layer and \(DWConv(\cdot)\) denotes a depth-wise convolutional layer. This two-step convolution ensures that the spatial and temporal features of EEG signals are effectively preserved and processed for the next steps.

Subsequently, we split the tensor \(E' \) along the channel dimension to obtain the query \(Q\), key \(K\) and value \(V\) components used for the self-attention mechanism~\cite{vaswani2017attention}.
\begin{equation}
    Q, K, V = Split(E') \in \mathbb{R}^{C \times 1 \times T}
\end{equation}
where \(Split(\cdot)\) denotes the operation of splitting \(E'\) along the channel dimension.

Then, we introduce a MTA to comprehensively extract latent long-short range information and construct expressive temporal features \(V' \) with richer contextual details. The purpose of MTA is to further refine the value \(V\) representation by capturing diverse temporal dependencies at multiple scales.
\begin{equation}
    V' = MTA(V) \in \mathbb{R}^{C \times 1 \times T}
\end{equation}
The detailed design and functionality of MTA will be elaborated upon in the following subsection.
 
Then, by employing the effective self-attention operation across all channels, our method extracts the key spatial distribution features and multi-scale temporal features. Meantime, we apply a \(1 \times 1\) convolutional layer to further improve the nonlinear expression ability and obtain more robust spatiotemporal features \(H\) for the next block. The process can be formalized as:
\begin{equation}
    H = Conv(Attention(Q, K, V'))
\end{equation}
\begin{equation}
    Attention(Q, K, V') = Softmax(\frac{QK^{T}}{t})V' 
\end{equation} 
where \(Conv(\cdot)\) represents a \(1 \times 1\) convolutional layer without bias. \(Softmax(\cdot)\) denotes the softmax function and \(t\) is a learnable scaling parameter. 

Finally, the EEG data \(H\) is passed into MGA for further processing, constructing the refined spatiotemporal features \(F\).
\begin{equation}
    F = MGA(H) \in \mathbb{R}^{1 \times C \times T}
\end{equation}
The MGA will be discussed in detail in the following subsection.

\subsection{Multi-scale Temporal Attention Block}
Previous studies have demonstrated that the use of temporal attention mechanisms can effectively capture the time-varying nature of EEG signals~\cite{stanet,DBPNet}. However, they usually overlook the importance of multi-scale information and fail to fully consider long-short range dependencies inherent in temporal features. To address this limitation, we propose MTA to fully leverage time-series information by incorporating multiple temporal scales.

Firstly, a spatial filter is used to reduce the channel dimension to prevent an excessive number of trainable parameters. Next, we use a \(1 \times 1\) convolutional layer to increase the channel again and split EEG signals into three parts \(X, Y, Z\) for operations of different scales. It can be formally described as follows:
\begin{equation}
    X, Y, Z = Split(UpConv(SpatialConv(V))) 
\end{equation}
where \(SpatialConv(\cdot)\) denotes a \(C \times 1\) convolutional layer. \(UpConv(\cdot)\) denotes a \(1 \times 1\) convolutional layer with 3 output channels. \(Split(\cdot)\) performs the split operation along the channel dimension.

Next, \(X\), \(Y\) and \(Z\) are passed through different multi-scale temporal layers to compute multiple attention values \(\alpha,\beta,\gamma\). 
\begin{equation}
    \{\alpha,\beta,\gamma\} = AAP(ELU(LN(Conv_i(\{X,Y,Z\})))) 
\end{equation}
where \(Conv_i(\cdot)\) refers to a convolutional layer with a kernel size of \(i\in \{2, 4, 6\}\). \(LN(\cdot)\) denotes the layer normalization technique~\cite{Ba2016LayerN}. \(ELU(\cdot)\) is the exponential linear unit (ELU) activation function~\cite{clevert2015fast} and \(AAP(\cdot)\) denotes an adaptive average pooling layer. 

Then, we multiply each attention value with \(X, Y, Z\) separately and add them up to obtain the final fused temporal features \(V' \) with multi-scale temporal patterns. Finally, it is put into a \(1 \times 1\) convolutional layer to recover the number of channels and reshaped for further processing in the next stage.
\begin{equation}
    V' = Conv(\alpha \odot X + \beta \odot Y + \gamma \odot Z) 
\end{equation}
where \(\odot\) denotes the element-wise multiplication and \(Conv(\cdot)\) represents a \(1 \times 1\) convolutional layer with \(C\) output channels. \(\alpha,\beta,\gamma\) are the learnable attention weight values.

\subsection{Multi-scale Global Attention Block}
Most previous studies usually overlook the dynamic relationship between spatial distribution features and time-series data of EEG signals. Enlightened by ~\cite{wang2024multiscaleattentionnetworksingle}, we propose MGA to further extract latent spatiotemporal dependencies within EEG signals from a global perspective.

Firstly, a \(1 \times 1\) convolutional layer is applied to increase the channel dimension of the input data, producing \(H' \). This is followed by cloning the data to facilitate the final multi-scale fusion. Then, we split the data into three distinct parts \(P, S, R\) for further processing.
\begin{equation}
    H' = UpConv(H) \in \mathbb{R}^{3 \times C \times T}  
\end{equation}
\begin{equation}
    P, S, R = Split(H') \in \mathbb{R}^{1 \times C \times T}  
\end{equation}
where \(UpConv(\cdot)\) denotes a \(1 \times 1\) convolutional layer that outputs 3 channels. \(Split(\cdot)\) refers to a split operation along the channel dimension.

Subsequently, we separately compute the spatiotemporal attention maps. To capture multi-level spatiotemporal dependencies, we utilize dilated convolutional layers with varying kernel sizes and dilation rates. Each attention map is multiplied with its corresponding original features to obtain local refined attention maps \(\delta,\varphi,\mu\).
\begin{equation}
    \{\delta,\varphi,\mu\} = DConv_i(\{P,S,R\}) \odot \{P,S,R\} 
\end{equation}
where \(DConv_i(\cdot)\) represents \(i\)-th dilated convolutional layer with varying kernel size (\(3 \times 3\), \(5 \times 5\) and \(7 \times 7\)). The dilation rate of each layer is determined by the length of the decision window.

Next, we concatenate the resulting feature maps to obtain a multi-scale global attention map. It is then multiplied with \(H' \) to comprehensively capture spatiotemporal dependencies. Finally, a \(1 \times 1\) convolutional layer is employed to restore its original shape. Simultaneously, a residual connection~\cite{DeepResidualLearningforImageRecognition} is introduced to enhance the performance and robustness of the block. It can be formulated as follows: 
\begin{equation}
    F = DwConv(H'\odot [\delta, \varphi, \mu]) + H
\end{equation}
where \([\cdot]\) denotes the concatenate operation, which combines multiple attention maps along the channel dimension. \(DwConv(\cdot)\) represents a \(1 \times 1\) convolutional layer with 1 output channel.

\subsection{Spatiotemporal Convolution Module}
Through the comprehensive extraction of spatiotemporal features performed by the previous module, we obtain expressive information from EEG signals. To further improve the performance of AAD, STC is introduced at the end of our model to aggregate the extracted features. In STC, we apply a temporal layer and a spatial layer, followed by an adaptive average pooling layer to further process and downsample the data.
\begin{equation}
    F' = ELU(BN(TemporalConv(F)))  
\end{equation}
\begin{equation}
    F'' = ELU(BN(SpatialConv(F'))) 
\end{equation}
\begin{equation}
    O = AdaptiveAvgPool(F'')) 
\end{equation}
where \(TemporalConv(\cdot)\) denotes a 2D convolutional layer with a \(1 \times 2\) kernel size, and \(SpatialConv(\cdot)\) represents a 2D convolutional layer with a \(C \times 1\) kernel size across all the channels. \(BN(\cdot)\) is the batch normalization layer~\cite{BN} and \(AdaptiveAvgPool(\cdot)\) denotes an adaptive average pooling layer.

\section{Experiments}
\subsection{Datasets}

In this paper, we conduct extensive experiments on three publicly available datasets, namely KUL~\cite{das2016effect,das2019auditory}, DTU~\cite{fuglsang2017noise,fuglsang2018eeg} and AVED~\cite{Fan2024MSFNetMF}, as shown in Table~\ref{dataset-table}. The KUL and DTU datasets are among the most commonly used for AAD with the audio-only scene. The AVED dataset, provided for the ISCSLP Chinese AAD Challenge 2024, includes both audio-only and audio-visual scenes.
\begin{enumerate}
    \renewcommand{\labelenumi}{\arabic{enumi})}
    \item \textbf{KUL Dataset:} In this dataset, 64-channel EEG data were collected from 16 normal-hearing subjects (8 males and 8 females) at a sampling rate of 8192 Hz. The stimuli comprised four Dutch short stories narrated by three male speakers from 90° to the left or right. Each subject completed 8 trials, with each trial lasting 6 minutes.
    
    \item \textbf{DTU Dataset:} In this dataset, 64-channel EEG data were collected from 20 normal-hearing subjects at a sampling rate of 512 Hz. The auditory stimuli consisted of Danish audiobooks narrated by a male and a female speaker. Each subject completed 60 trials, with each trial lasting 50 seconds.
    
    \item \textbf{AVED Dataset:} In this dataset, 32-channel EEG data were collected from 20 normal-hearing subjects (14 males and 6 females) at a sampling rate of 1 kHz. The subjects were divided into two groups of 10. One group underwent audio-only experiments, and the other group underwent audio-video ones. All auditory stimuli were derived from 16 stories selected from a collection of Chinese short stories narrated by a male and a female speaker. Each subject completed 16 trials, with each trial lasting 152 seconds.
\end{enumerate}

\begin{table}[ht]
    \centering
    \scalebox{0.87}{
    \begin{tabular}{ccccc}
        \toprule[1.3pt]
        \textbf{Dataset} & \textbf{Subjects} & \textbf{Scene} & \parbox{2cm}{\centering \textbf{Duration (minutes)} } & \textbf{Language} \\
        \midrule
        KUL & 16 & audio-only  & 48 & Dutch\\
        \midrule
        DTU & 18 & audio-only  & 10 & Danish\\
        \midrule
        \multirow{2.5}{*}{AVED} & 10 & audio-only & 40 & Mandarin \\
        \cmidrule(r){2-5}
         & 10 & audio-visual & 40 & Mandarin\\
        \bottomrule[1.3pt]
    \end{tabular}
    }
    \caption{Details of three datasets used in the experiments.}
    \label{dataset-table}
\end{table}

\begin{table*}[ht]
    \centering
    \begin{tabular}{cccccc}
        \toprule[1.3pt]
        \multirow{2.5}{*}{\textbf{Decision Window}} & \multirow{2.5}{*}{\textbf{Model}} & \multicolumn{4}{c}{\textbf{Dataset}} \\
        \cmidrule(r){3-6}
        & & \textbf{KUL} & \textbf{DTU} & \textbf{AVED (AO)} & \textbf{AVED (AV)} \\
        \midrule
        \multirow{6}{*}{0.1-second} & SSF-CNN\textsuperscript{*} ~\cite{cai2021low}
        & 76.3 \(\pm\) 8.47 & 62.5 \(\pm\) 3.40 & 53.3 \(\pm\) 1.91 & 54.2 \(\pm\) 2.00 \\
        & MBSSFCC\textsuperscript{*} ~\cite{mbssf} & 79.0 \(\pm\) 7.34 & 66.9 \(\pm\) 5.00  & 57.6 \(\pm\) 2.87 & 58.9 \(\pm\) 2.60 \\  
        & EEG-Graph Net ~\cite{cai2023brain} & - & 72.5 \(\pm\) 7.41  & - & - \\
        & DBPNet\textsuperscript{*} ~\cite{DBPNet}& 85.3 \(\pm\) 6.22 & 74.0 \(\pm\) 5.20 & 53.6  \(\pm\) 2.93 & 55.7  \(\pm\) 2.45 \\
        & DARNet\textsuperscript{*} ~\cite{yan2024darnet}& 89.2 \(\pm\) 5.50 & 74.6 \(\pm\) 6.09 & 51.3 \(\pm\) 3.50 & 50.3 \(\pm\) 0.60 \\
        & \textbf{MHANet(ours)} & \textbf{95.6} \(\bm{\pm}\) \textbf{4.83} & \textbf{75.5} \(\bm{\pm}\) \textbf{5.68} & \textbf{67.9} \(\bm{\pm}\) \textbf{2.10} & \textbf{67.4} \(\bm{\pm}\) \textbf{3.24} \\
        \midrule
        \multirow{8}{*}{1-second} & SSF-CNN\textsuperscript{*} ~\cite{cai2021low}
        & 84.4 \(\pm\) 8.67 & 69.8 \(\pm\) 5.12 & 57.1 \(\pm\) 3.54 & 59.2 \(\pm\) 5.13 \\
        & MBSSFCC\textsuperscript{*} ~\cite{mbssf} & 86.5 \(\pm\) 7.16 & 75.6 \(\pm\) 6.55  & 70.5 \(\pm\) 3.92 & 69.5 \(\pm\) 5.77 \\  
        & DGSD ~\cite{fan2024dgsd} & 90.3 \(\pm\) 7.29 & 79.6 \(\pm\) 6.76 & - & -  \\
        & EEG-Graph Net ~\cite{cai2023brain} & - & 78.7 \(\pm\) 6.47  & - & - \\
        & DenseNet-3D ~\cite{xu2024densenet} & 94.3 \(\pm\) 4.3 & - & - & - \\
        & DBPNet\textsuperscript{*} ~\cite{DBPNet}& 94.4 \(\pm\) 4.62 & 79.8 \(\pm\) 6.91 & 58.7  \(\pm\) 3.60 & 62.0  \(\pm\) 4.92 \\
        & DARNet\textsuperscript{*} ~\cite{yan2024darnet}& 94.8 \(\pm\) 4.53 & 80.1 \(\pm\) 6.85 & 80.6 \(\pm\) 15.69 & 83.1 \(\pm\) 11.64 \\
        & \textbf{MHANet(ours)} & \textbf{95.8} \(\bm{\pm}\) \textbf{4.29} & \textbf{82.2} \(\bm{\pm}\) \textbf{8.13} & \textbf{87.1} \(\bm{\pm}\) \textbf{4.48} & \textbf{86.0} \(\bm{\pm}\) \textbf{5.32} \\
        \midrule
        \multirow{8}{*}{2-second} & SSF-CNN\textsuperscript{*} ~\cite{cai2021low}
        & 87.8 \(\pm\) 7.87 & 73.3 \(\pm\) 6.21 & 59.8 \(\pm\) 4.72 & 63.4 \(\pm\) 5.13 \\
        & MBSSFCC\textsuperscript{*} ~\cite{mbssf} & 89.5 \(\pm\) 6.74 & 78.7 \(\pm\) 6.75  & 76.2 \(\pm\) 3.64 & 74.3 \(\pm\) 7.04 \\  
        & DGSD ~\cite{fan2024dgsd} & 93.3 \(\pm\) 6.53 & 82.4 \(\pm\) 6.86 & - & -  \\
        & EEG-Graph Net ~\cite{cai2023brain} & - & 79.4 \(\pm\) 7.16  & - & - \\
        & DenseNet-3D ~\cite{xu2024densenet} & 95.9 \(\pm\) 4.3 & - & - & - \\
        & DBPNet\textsuperscript{*} ~\cite{DBPNet}& 95.3 \(\pm\) 3.50 & 80.2 \(\pm\) 6.79 & 62.2  \(\pm\) 6.27 & 63.3  \(\pm\) 4.56 \\
        & DARNet\textsuperscript{*} ~\cite{yan2024darnet}& 95.5 \(\pm\) 4.89 & 81.2 \(\pm\) 6.34 & 91.3 \(\pm\) 2.73 & 87.6 \(\pm\) 13.19 \\
        & \textbf{MHANet(ours)} & \textbf{96.6} \(\bm{\pm}\) \textbf{3.67}  & \textbf{83.0} \(\bm{\pm}\) \textbf{7.14} & \textbf{92.9} \(\bm{\pm}\) \textbf{3.93} & \textbf{92.0} \(\bm{\pm}\) \textbf{3.84} \\
        \bottomrule[1.3pt]
    \end{tabular}
    \caption{Auditory attention detection accuracy (\%) comparison on the KUL, DTU, and AVED datasets. The KUL and DTU datasets consist of the audio-only scene. The AVED dataset includes both audio-only (AO) and audio-visual (AV) scenes. – indicates that no corresponding experiments are conducted or no results are provided in the respective paper. The results of the baseline models marked with * have been reproduced.}
    \label{compare-table}
\end{table*}

\subsection{Data Processing}
To ensure a fair comparison of the performance of our MHANet, specific preprocessing steps are applied to each dataset. For the KUL dataset, EEG data are initially re-referenced to the average response of mastoid electrodes, followed by bandpass filtering between 0.1 Hz and 50 Hz. The data are then down-sampled to 128 Hz. For the DTU dataset, EEG data are processed to remove 50 Hz linear noise and its harmonics. Eye artefacts are removed through joint decorrelation, and then the data are re-referenced to the average response of mastoid electrodes. Finally, the data are down-sampled to 64 Hz. For the AVED dataset, a notch filter is applied first to eliminate powerline interference at 50 Hz. Next, a finite impulse response (FIR) filter is used for high-pass and low-pass filtering to remove the noise. Subsequently, the EEG data are downsampled to 128 Hz, followed by independent component analysis (ICA) for further noise removal. Finally, a re-referencing process is performed across all EEG channels to ensure consistency and comparability.

\subsection{Implementation Details}
In previous research on AAD, classification accuracy has been utilized as the standard metric for evaluating model performance. Following this convention, we assess our proposed MHANet using the KUL, DTU, and AVED datasets. To illustrate implementation details, including training settings and network configuration, we use the KUL dataset as an example, with a 1-second decision window.

The dataset is initially divided into training, validation, and test sets in a ratio of 8:1:1. For each subject in the KUL dataset, we allocate 4,600 decision windows for training, 576 for validation, and 576 for testing. The training process uses a batch size of 32, with a maximum of 100 epochs. An early stopping strategy is employed, halting training if there is no decrease in the validation set's loss function value for 15 consecutive epochs. The model is trained using the AdamW optimizer with a learning rate of 5e-3 and weight decay of 3e-4.

Initially, we employ the common spatial patterns (CSP) algorithm~\cite{ramoser2000optimal,blankertz2007optimizing} to extract raw features from the EEG signals and rearrange them into \(E \). Then, through MHA, we get the refined spatiotemporal features \(F\). It is then sent to STC. After convolutional layers and global average pooling, we obtain \(O \). Finally, the final binary AAD classification result \(p \) is achieved through a fully connected layer (input: 5, output: 2). 
\section{Results on AAD}

\subsection{Performance of MHANet}
To comprehensively evaluate our proposed MHANet, we assess its AAD performance under different decision windows and compare our model with other outstanding methods, as shown in Table~\ref{compare-table}. For open-access models, we replicate their architectures, while results from other studies are cited accordingly. 

Our MHANet demonstrates significant improvement and superior performance on KUL, DTU, and AVED datasets. On the DTU dataset, the accuracies are 75.5\% (SD: 5.68\%), 82.2\% (SD: 8.13\%), 83.0\% (SD: 7.14\%) under the 0.1-second, 1-second and 2-second decision windows, respectively. Similar accuracy trends are observed on the KUL and AVED datasets at different lengths of the decision window.

Overall, our model's performance decreases as the decision window length shortens. However, we observe that on the KUL dataset, the performance under 0.1-second is nearly as good as under 1-second. This could be because, under a 0.1-second decision window, the number of windows significantly increases, providing the model with more training samples, which aids in its learning. At the same time, the performance on the DTU dataset is generally lower compared to the KUL dataset. This could be attributed to differences in stimulus source locations, the number of speakers, their genders, and variations in data processing.

\subsection{Ablation Experiment}
To thoroughly analyze our model, we conduct extensive ablation experiments by removing the CA mechanism, MTA, MGA, both MTA and CA, and STC. All experiments are conducted under the same conditions as the previous settings. The results of these ablation experiments are presented in Table~\ref{ablation-table}.

Experimental results show that on the DTU dataset, removing CA from MHANet leads to a decrease in average accuracy by 8.6\% under the 1-second decision window. Removing the MTA also causes a similar accuracy drop: 3.7\% for the 1-second decision window. When the MGA is removed, the accuracy decreases by  0.5\% for the 1-second decision window. After removing both MTA and CA, the average accuracy decreases by 11.2\% under the 1-second decision window. The removal of STC results in accuracy drops of 2.5\% for the 1-second decision window. Similar trends of decreased accuracy are observed on the KUL and AVED datasets and at different lengths of the decision window after removing the aforementioned modules or blocks.

Overall, the complete MHANet demonstrates the best performance compared to versions with individual modules or blocks removed. On the DTU dataset with a 1-second decision window, MHANet outperforms the version without CA by 8.6\%, highlighting the importance of focusing on spatial distribution features within EEG signals. It also achieves a 3.7\% improvement over the version without MTA, indicating the effectiveness of capturing multi-scale temporal patterns. Additionally, MHANet shows a 0.5\% improvement compared to the version without MGA, demonstrating the value of considering spatiotemporal dependencies from a global perspective. Finally, it outperforms the version without STC by 2.5\%, emphasizing the necessity of integrating and refining spatiotemporal features of EEG signals from different dimensions.
\begin{table*}[ht]
    \centering
    \begin{tabular}{cccccc}
        \toprule[1.3pt]
        \multirow{2.5}{*}{\textbf{Decision Window}} & \multirow{2.5}{*}{\textbf{Model}} & \multicolumn{4}{c}{\textbf{Dataset}} \\
        \cmidrule(r){3-6}
        & & \textbf{KUL} & \textbf{DTU} & \textbf{AVED (AO)} & \textbf{AVED (AV)} \\
        \midrule
        \multirow{6}{*}{0.1-second} & w/o CA
        & 80.2 \(\pm\) 12.03 & 66.1 \(\pm\) 6.69 & 49.5 \(\pm\) 1.50 & 50.0 \(\pm\) 0.15 \\
        & w/o MTA & 95.3 \(\pm\) 3.81 & 72.8 \(\pm\) 5.89  & 55.9 \(\pm\) 6.35 & 55.9 \(\pm\) 5.91 \\  
        & w/o MGA & 95.2 \(\pm\) 4.87 & 74.6 \(\pm\) 5.85 & 67.4 \(\pm\) 2.28 & 67.1 \(\pm\) 3.22 \\
        & w/o MTA and CA & 89.1 \(\pm\) 5.89 & 71.4 \(\pm\) 8.12 & 50.6  \(\pm\) 1.82 & 51.3  \(\pm\) 2.69 \\
        & w/o STC & 94.4 \(\pm\) 5.72 & 74.7 \(\pm\) 5.71 & 67.7 \(\pm\) 2.38 & 67.3 \(\pm\) 3.44 \\
        & \textbf{MHANet(ours)} & \textbf{95.6} \(\bm{\pm}\) \textbf{4.83} & \textbf{75.5} \(\bm{\pm}\) \textbf{5.68} & \textbf{67.9} \(\bm{\pm}\) \textbf{2.10} & \textbf{67.4} \(\bm{\pm}\) \textbf{3.24} \\
        \midrule
        \multirow{6}{*}{1-second} & w/o CA
        & 82.8 \(\pm\) 12.10 & 73.6 \(\pm\) 9.91 & 48.4 \(\pm\) 3.65 & 50.6 \(\pm\) 3.87 \\
        & w/o MTA & 95.5 \(\pm\) 4.69 & 78.5 \(\pm\) 8.62  & 48.1 \(\pm\) 4.72 & 68.4 \(\pm\) 13.18 \\  
        & w/o MGA & 95.5 \(\pm\) 4.44 & 81.7 \(\pm\) 8.74 & 86.9 \(\pm\) 3.96 & 85.3 \(\pm\) 4.94 \\
        & w/o MTA and CA & 93.1 \(\pm\) 4.86 & 71.0 \(\pm\) 9.60  & 49.6 \(\pm\) 1.75 & 50.1 \(\pm\) 1.69 \\
        & w/o STC & 95.6 \(\pm\) 4.01 & 79.7 \(\pm\) 8.32 & 87.0 \(\pm\) 3.43 & 84.4 \(\pm\) 4.81 \\
        & \textbf{MHANet(ours)} & \textbf{95.8} \(\bm{\pm}\) \textbf{4.29} & \textbf{82.2} \(\bm{\pm}\) \textbf{8.13} & \textbf{87.1} \(\bm{\pm}\) \textbf{4.48} & \textbf{86.0} \(\bm{\pm}\) \textbf{5.32} \\
        \midrule
        \multirow{6}{*}{2-second} & w/o CA
        & 81.5 \(\pm\) 11.99 & 69.9 \(\pm\) 11.65 & 50.7 \(\pm\) 2.22 & 50.2 \(\pm\) 0.67 \\
        & w/o MTA & 96.1 \(\pm\) 4.45 & 78.9 \(\pm\) 7.64  & 60.7 \(\pm\) 17.86 & 60.1 \(\pm\) 15.92 \\  
        & w/o MGA & 95.9 \(\pm\) 4.51 & 82.4 \(\pm\) 7.25 & 91.4 \(\pm\) 2.60 & 91.3 \(\pm\) 5.35  \\
        & w/o MTA and CA & 94.6 \(\pm\) 4.23 & 72.8 \(\pm\) 8.33  & 49.7 \(\pm\) 0.94 & 49.6 \(\pm\) 1.07 \\
        & w/o STC & 95.7 \(\pm\) 4.60 & 79.7 \(\pm\) 8.46 & 89.8 \(\pm\) 5.68 & 88.6 \(\pm\) 4.26 \\
        & \textbf{MHANet(ours)} & \textbf{96.6} \(\bm{\pm}\) \textbf{3.67}  & \textbf{83.0} \(\bm{\pm}\) \textbf{7.14} & \textbf{92.9} \(\bm{\pm}\) \textbf{3.93} & \textbf{92.0} \(\bm{\pm}\) \textbf{3.84} \\
        \bottomrule[1.3pt]
    \end{tabular}
    \caption{Ablation Study on KUL, DTU, and AVED dataset. The KUL and DTU datasets consist of the audio-only scene. The AVED dataset includes both audio-only (AO) and audio-visual (AV) scenes. CA represents the channel attention. MTA denotes the multi-scale temporal attention block. MGA represents the multi-scale global attention block. STC denotes the spatiotemporal convolution module.}
    \label{ablation-table}
\end{table*}

\section{Analysis and Discussion}
\subsection{Comparison with the SOTA models}
We compare the performance of our MHANet with other advanced AAD models, as presented in Table~\ref{compare-table}. The results demonstrate that MHANet achieves significant improvements over the current SOTA methods.

On the DTU dataset, our MHANet achieves relative improvements of 13.0\%, 8.6\%, 3.0\%, 1.5\%, and 0.9\% under the 0.1-second decision window compared to the SSF-CNN, MBSSFCC, EEG-Graph Net, DBPNet and DARNet models, respectively. For the 1-second decision window, MHANet achieves relative improvements of 12.4\%, 6.6\%, 2.6\%, 3.5\%, 2.4\%, and 2.1\% compared to the SSF-CNN, MBSSFCC, DGSD, EEG-Graph Net, DBPNet, and DARNet models. Similarly, the relative improvements under the 2-second decision window are 9.7\%, 4.3\%, 0.6\%, 3.6\%, 2.8\%, and 1.8\%, respectively. 

On both the KUL and AVED datasets, MHANet achieves comparable improvements over other models. Notably, on the KUL dataset, MHANet achieves a 6.4\% relative improvement under the 0.1-second decision window compared to the current SOTA method. At the same time, on the AVED dataset, the accuracies of other models are generally lower under the 0.1-second decision window, while our model shows a significant improvement. These all suggest that our model has an advantage in tasks that require adaptation to extremely short time windows. 

Overall, the outstanding performance of MHANet across different datasets and decision windows highlights its strong real-world practicality and applicability, making it a promising solution for realistic hearing aids. 

\subsection{Ablation Analysis}
As shown in Table~\ref{ablation-table}, the results demonstrate that removing the CA, MTA, MGA, MTA and CA or STC leads to performance degradation. This highlights the effectiveness of each block and module in contributing to the advanced performance of our MHANet.

\subsubsection{Effectiveness of CA}
The distribution of EEG channels reflects the spatial relationships between them. Therefore, it is crucial to exploit the spatial coherence across different EEG channels~\cite{Electroencephalography-BasedAuditoryAttentionDecoding:TowardNeurosteeredHearingDevices}. The incorporation of the channel attention mechanism enables our model to focus on key spatial information within EEG signals, thus capturing the activities of different brain regions, which is essential for effectively analyzing EEG signals.

\subsubsection{Effectiveness of MTA}
Our MTA effectively captures multi-scale temporal information and generates more expressive and robust temporal features through convolutional attention at different scales. Therefore, it further improves the model's ability to understand latent temporal context at different levels and identify long-short range temporal patterns within EEG signals.

\subsubsection{Effectiveness of MGA}
Our MGA fully leverages the spatiotemporal feature maps to extract more valuable spatial distribution features and temporal patterns at multiple scales. By treating the EEG data as Euclidean data, it can capture spatiotemporal dependencies of EEG signals from a global perspective. The results show that our MGA further enhances performance to some extent, demonstrating the effectiveness of simultaneous spatiotemporal consideration.

\subsubsection{Effectiveness of STC}
The STC effectively aggregates spatiotemporal features along the temporal and channel dimensions through convolutional operations, respectively, resulting in more robust features with reduced noise and outliers. Meanwhile, the integration enhances our model's ability to understand brain activity, further improving its overall robustness and generalization ability.

\subsection{Comparison of Computational Cost}
We compare the training parameter counts of our MHANet, SSF-CNN ~\cite{cai2021low}, MBSSFCC ~\cite{mbssf}, DBPNet ~\cite{DBPNet} and DARNet ~\cite{yan2024darnet}. The results are shown in Table~\ref{parameters-table}. The parameter count of MHANet is 209.5 times lower than that of SSF-CNN, 4194.5 times lower than MBSSFCC, 44.5 times lower than DBPNet, and 3 times lower than that of DARNet. MHANet showcases excellent parameter efficiency by achieving competitive performance with a significantly reduced number of parameters. This suggests our model's strong real-world applicability and practicality, making it well-suited for use in real-world low-resource hearing aids.
\begin{table}
    \centering
    \begin{tabular}{cc}
        \toprule[1.3pt]
        Model & Trainable Parameters\\
        \midrule
         SSF-CNN ~\cite{cai2021low} & 4.21 M \\
         MBSSFCC ~\cite{mbssf} & 83.91 M \\
         DBPNet ~\cite{DBPNet} & 0.91 M \\
         DARNet ~\cite{yan2024darnet} & 0.08 M \\
         \textbf{MHANet (ours)} & \textbf{0.02 M} \\
        \bottomrule[1.3pt]
    \end{tabular}
    \caption{The training parameter counts of our MHANet and four open-source models. "M" denotes a million.}
    \label{parameters-table}
\end{table}

\section{Conclusion}
This paper proposes MHANet, a novel multi-scale hybrid attention network, to address the oversight of multi-scale spatiotemporal dependencies in AAD. We introduce MHA to capture long-short range spatiotemporal dependencies of EEG signals. The MHA combines CA with MTA to construct comprehensive and robust EEG features. Then, the MGA extracts the dynamic relationship between key spatial distribution features and temporal patterns. Subsequently, STC aggregates expressive EEG signals, improving our model's overall robustness and generalization ability. We evaluate the performance of the proposed MHANet on three datasets: KUL, DTU, and AVED, which demonstrate that MHANet achieves SOTA performance with fewer trainable parameters across all three datasets. This highlights its strong real-world practicality and applicability for realistic hearing aids. For future research, we plan to incorporate time-frequency analysis to further explore the spatiotemporal dependencies within EEG signals and enhance the performance of AAD.

\section*{Acknowledgments}
This work is supported by the {STI 2030—Major Projects (No. 2021ZD0201500)}, the National Natural Science Foundation of China (NSFC) (No.62201002, 6247077204), Excellent Youth Foundation of Anhui Scientific Committee (No. 2408085Y034), Distinguished Youth Foundation of Anhui Scientific Committee (No. 2208085J05), Special Fund for Key Program of Science and Technology of Anhui Province (No. 202203a07020008), Cloud Ginger XR-1.

\section*{Contribution Statement}
Lu Li and Cunhang Fan contributed equally to this work.
\appendix

\bibliographystyle{named}
\bibliography{ijcai25}

\end{document}